\documentclass[runningheads,a4paper]{llncs}
\usepackage{makeidx}  
\usepackage{amssymb}
\usepackage{amsmath}
\usepackage{amsfonts}
\usepackage{graphicx}
\usepackage{multirow}
\usepackage{cite}
\usepackage{makecell}

\usepackage{url}

\begin{document}

\mainmatter   

\title{Bayesian Regression Approach  for  \\ Building and Stacking 
        Predictive Models \\ in Time Series Analytics}
\titlerunning{Bayesian Regression Approach  for  Building And Stacking 
        Predictive Models}
\author{Bohdan M. Pavlyshenko}

\institute{Ivan Franko National University of Lviv \\
Lviv, Ukraine \\
	b.pavlyshenko@gmail.com,  www.linkedin.com/in/bpavlyshenko/}
\maketitle  
\begin{abstract}
The paper describes the use of Bayesian regression for  building time series models and  stacking   different predictive models for time series. Using Bayesian regression for  time series modeling with nonlinear trend was analyzed. 
This  approach makes it possible to estimate an uncertainty of time series prediction and  calculate value at risk  characteristics. 
A hierarchical model for time series using Bayesian regression has been considered. 
In this approach, one set of parameters is the same for all data samples, other parameters can be different for different groups of data samples. Such an approach 
 allows  using this  model in the case of short historical data for specified time series, e.g. in the case of new stores or new products in the sales prediction problem. 
In the study of predictive models stacking, the models  ARIMA, Neural Network, Random Forest, Extra Tree were used for the prediction on the first level of model ensemble. 
On the second  level, time series predictions of these models on the validation set were used for stacking by Bayesian regression. This approach gives distributions for regression coefficients of these models. It makes it possible to estimate  the uncertainty contributed by each model  to stacking result.  
The information about these distributions allows us to select an optimal set of stacking models, taking into account the domain knowledge.  The probabilistic approach for stacking predictive 
models allows us to make risk assessment for the predictions that are important in a decision-making process. 

\end{abstract}

\section{Introduction}
Time series analytics is an important part of modern data science. 
The examples of different time series approaches are  in ~\cite{chatfield2000time, brockwell2002introduction, box2015time,
hyndman2018forecasting,
tsay2005analysis, wei2006time, hyndman2007automatic}.
In ~\cite{wolpert1992stacked, rokach2010ensemble, sagi2018ensemble, gomes2017survey, dietterich2000ensemble, rokach2005ensemble},   
 a range of ensemble-based methods for classification problems is considered. 
In ~\cite{papacharalampous2018univariate}, the   authors studied 
a lagged variable selection, hyperparameter optimization, as well as compared classical and machine learning 
based algorithms for time series. 
Sales prediction is an important part of modern business intelligence ~\cite{mentzer2004sales,efendigil2009decision,zhang2004neural}. 
 Sales can be regarded as a time series.
However, time series approaches have some limitations for sales forecasting. For instance, to study seasonality, historical data  for large time period are required. 
But it frequently happens that there are no historical data for a target variable, e.g. when a new product is launched. 
 Furthermore, we must take into consideration many exogenous factors  influencing sales.   
We can consider time series forecasting as a regression problem in many cases, 
especially for sales time series. 
 In ~\cite{pavlyshenko2016linear}, we considered linear models, machine learning and probabilistic models for time series modeling. 
In~\cite{pavlyshenko2016machine}, we regarded the logistic regression with Bayesian inference for analysing  manufacturing failures. 
In ~\cite{pavlyshenko2019machine}, we study the use of machine learning models for sales predictive analytics. We researched the main approaches and case studies of implementing machine learning to sales forecasting. 
The effect of machine learning generalization has been studied.  This effect can be used for predicting sales in case of a small number of historical data for specific sales time series in case when a new product or store is launched ~\cite{pavlyshenko2019machine}. A stacking approach for building ensemble of single models using Lasso regression has also been studied.  
The obtained results show that using stacking techniques, we can improve the efficiency of predictive models for sales time series forecasting ~\cite{pavlyshenko2019machine}.
 
 In this paper,  we consider the use of Bayesian regression for building time series predictive models and  for stacking time series predictive models on the second level of the predictive model which is the ensemble of the models of the first level.

\section{Bayesian  Regression Approach}
Probabilistic regression models can be based on  Bayesian theorem ~\cite{kruschke2014doing,gelman2013bayesian, carpenter2017stan}. This approach allows us to receive a posterior distribution of model parameters using conditional likelihood and prior distribution. 
Probabilistic approach is more natural for stochastic variables such as sales time series. The difference between Bayesian approach and conventional Ordinary Least Squares (OLS) method is that in the Bayesian approach, uncertainty comes from parameters of model, as opposed to OLS method where the parameters are constant and uncertainty comes from data. In the Bayesian inference, we can use informative prior distributions which can be set up by an expert. So, the result can be considered as a compromise between 
historical data and expert opinion. It is important in the cases when we have a small number of historical data. In the Bayesian model, we can consider the target variable with non Gaussian  distribution, e.g. Student's t-distribution. 
Probabilistic approach enables us an ability to receive probability density function for the target variable. Having such function, we can make risk assessment and calculate value at risk (VaR) which is 5\% quantile. 
For solving Bayesian models, the numerical Monte-Carlo methods are used. Gibbs and Hamiltonian sampling are the popular methods of finding posterior distributions for the parameters of probabilistic model
~\cite{kruschke2014doing,gelman2013bayesian, carpenter2017stan}.

Bayesian inference makes it possible to do nonlinear regression.
If we need to fit trend with saturation, we can use logistic curve model for nonlinear trend and find its parameters using Bayesian inference. 
Let us consider the case of nonlinear regression for time series which have the trend with saturation. 
For modeling we consider sales time series. The model can be described as:
\begin{equation}
\begin{split}
& log(Sales) \sim \mathcal{N}(\mu_{Sales}, \, \sigma^2) \\
& \mu_{Sales}=\frac{a}{1+exp(b t +c)} +\beta_{Promo}Promo+ \\
&  \beta_{Time}Time+ \sum_{j} \beta^{wd}_{j}WeekDay_{j}, \\
\end{split}
\label{bayes_ts_eq1}
\end{equation}
where $WeekDay_{j}$ - are binary variables, which is 1 in the case of sales on an appropriate day of week  and 0 otherwise.   
We  can also build hierarchical models using Bayesian inference. 
In this approach, one set of parameters is the same for all data samples, other parameters can be different for different groups of data samples. In the case of sales data, we can consider trends, promo impact and seasonality as the same for all stores. But the impact of a specific store on sales can be described by an intersect parameter, so this coefficient for this variable will be different for different stores. 
The hierarchical model can be described as:
\begin{equation}
\begin{split}
& Sales \sim \mathcal{N}(\mu_{Sales}, \, \sigma^2) \\
& \mu_{Sales}=\alpha(Store)+\beta_{Promo}Promo+ \\
&  \beta_{Time}Time+ \sum_{j} \beta^{wd}_{j}WeekDay_{j}, \\
\end{split}
\label{eq_hierarchical_model_1}
\end{equation}
where intersect parameters $\alpha(Store)$ are different for different stores. 

Predictive models can be combined into ensemble model using stacking approach
~\cite{wolpert1992stacked, rokach2010ensemble, sagi2018ensemble, gomes2017survey, dietterich2000ensemble, rokach2005ensemble}. 
In this approach, prediction results of predictive models on the validation set are 
treated  as covariates for stacking regression. These predictive models are considered as a first level of predictive model ensemble. Stacking model forms the second level of model ensemble.  Using Bayesian inference for stacking regression gives distributions for stacking regression coefficients. It enables us to estimate 
the uncertainty of the first level predictive models.
As predictive models for first level of ensemble, we used the following 
models: ARIMA', 'ExtraTree', 'RandomForest', 'Lasso', 'NeuralNetowrk'. 
The use of these models for stacking by Lasso regression was described in 
~\cite{pavlyshenko2019machine}. 
For stacking, we have chosen the  robust regression with Student's t-distribution for the target variable
as 
 \begin{equation}
y \sim  Student_t (\nu, \mu, \sigma) 
\label{eq_1}
\end{equation}
where 
\begin{equation}
 \mu = \alpha + \sum_{i}{\beta_{i}x_{i} }, 
 \label{eq_2}
\end{equation}
$\nu$ is a distribution parameter, called as degrees of freedom, 
$i$ is an index of the predictive model in the stacking regression, $i \in$ \{ 'ARIMA', 'ExtraTree', 'RandomForest', 'Lasso', 'NeuralNetowrk' \}.
\section{Numerical Modeling}
\begin{figure}
\centering
\includegraphics[width=0.85\linewidth]{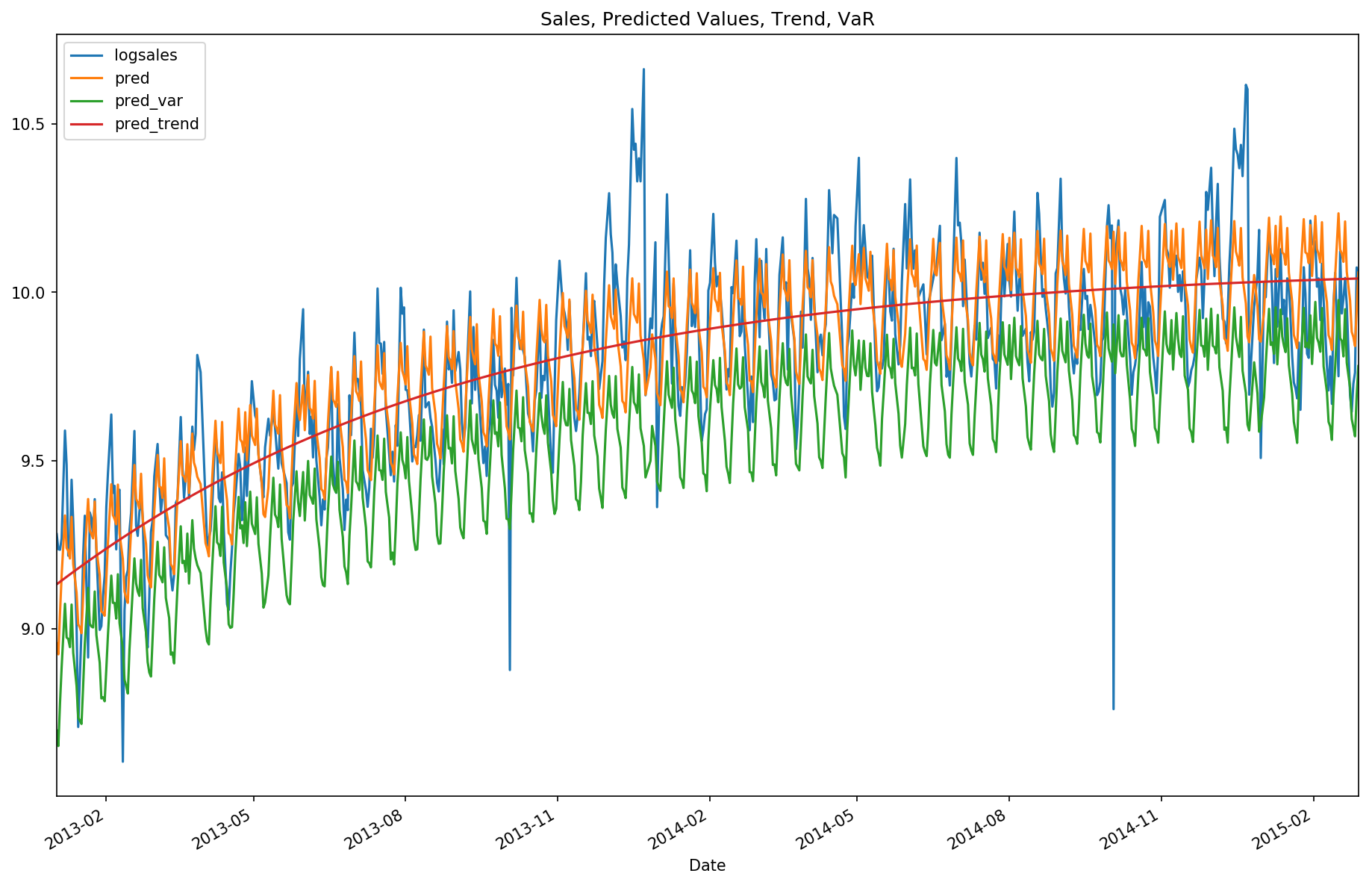}
\caption{Time series and trend forecasting}
\label{bayesian_ts_fig1}
\end{figure}

\begin{figure}
\centering
\includegraphics[width=0.55\linewidth]{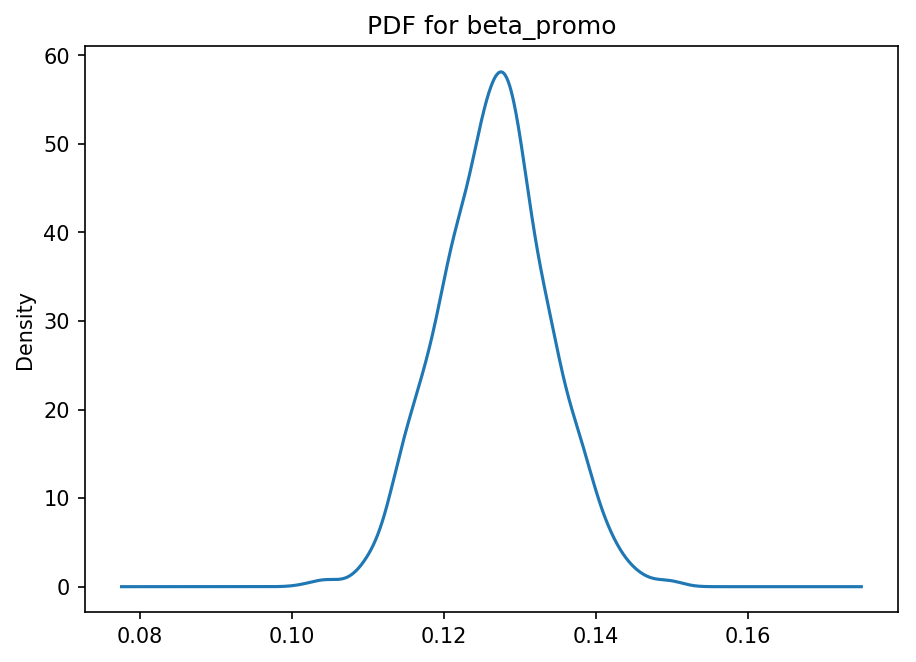}
\caption{Probability density function of regression coefficients for Promo factor}
\label{bayesian_ts_fig2}
\end{figure}

\begin{figure}
\centering
\includegraphics[width=0.75\linewidth]{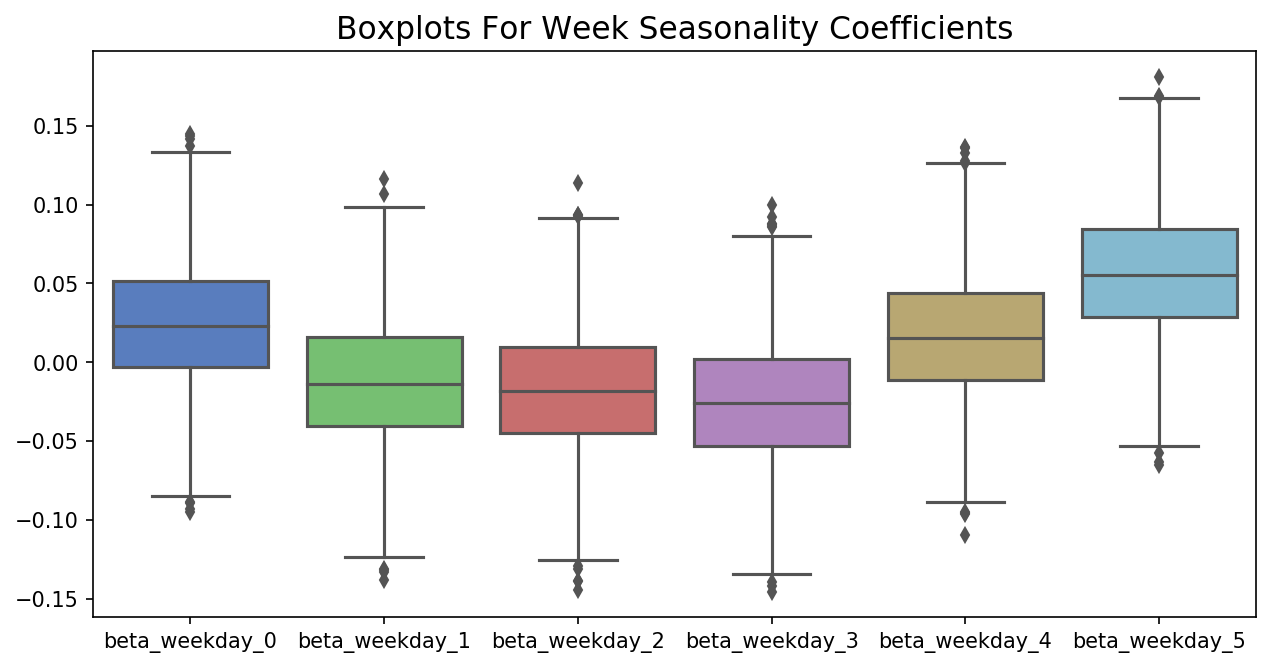}
\caption{Box plots for probability density function of seasonality coefficients}
\label{bayesian_ts_fig3}
\end{figure}

The data for our analysis are based on store sales historical data from  the ``Rossmann Store Sales'' Kaggle competition ~\cite{rossmanstorekaggle}.  For Bayesian regression, we used Stan platform for statistical modeling 
~\cite{carpenter2017stan}. The analysis was conducted in Jupyter Notebook environment  using Python programming language and 
the following  main Python packages \textit{pandas, sklearn, pystan,  numpy, scipy,statsmodels,  keras, matplotlib, seaborn}.  
For numerical analysis we modeled time series with multiplicative trend with saturation. 
Figure~\ref{bayesian_ts_fig1} shows results of this modeling where mean values and Value at Risk (VaR) characteristics are given. VaR was calculated as 5\% percentile. Figure~\ref{bayesian_ts_fig2} shows 
probability density function of regression coefficients for Promo factor. 
Figure~\ref{bayesian_ts_fig3} shows box plots for probability density function of seasonality coefficients. 

\begin{figure}
\centering
\includegraphics[width=0.65\linewidth]{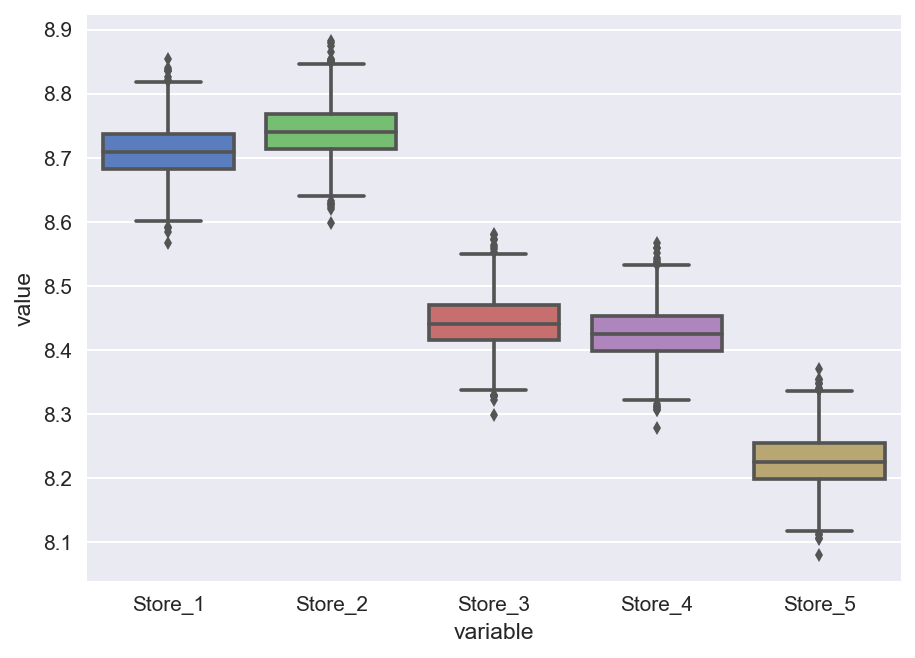}
\caption{Boxplots for  probability density functions  of intersect parameters}
\label{bayesian_ts_fig4}
\end{figure}

\begin{figure}
\centering{\includegraphics[width=0.8\linewidth]{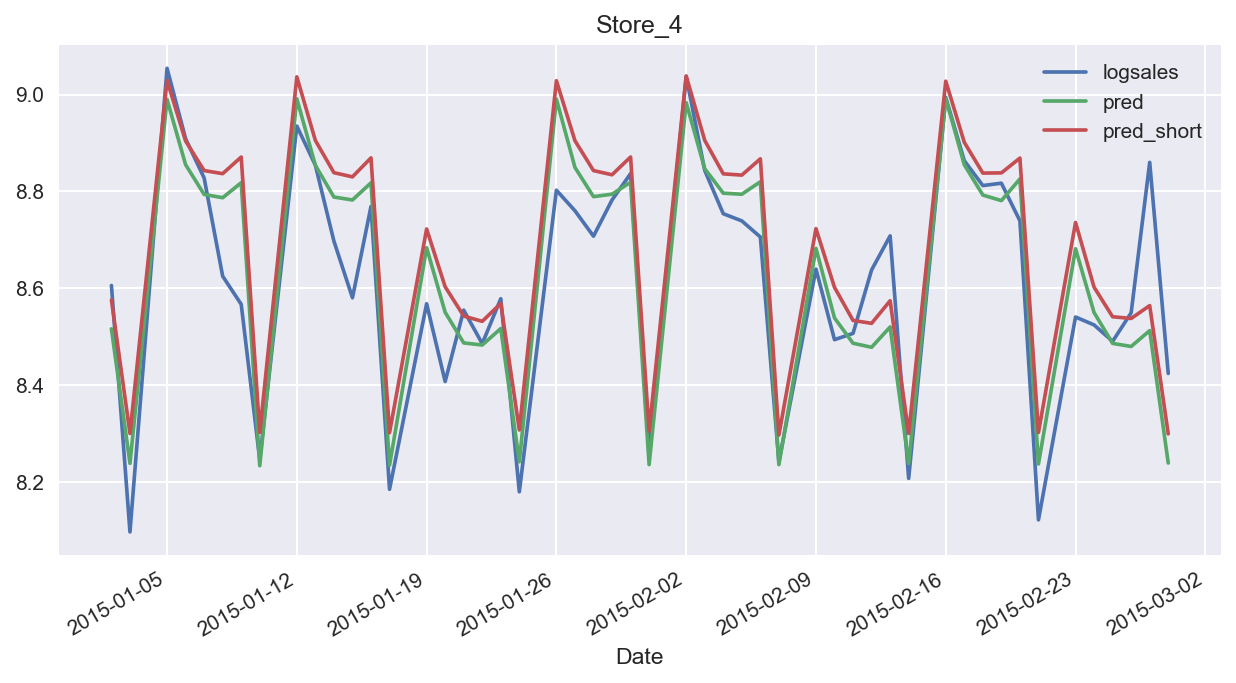}}
\caption{ Forecasting on the validation set for specified time series in cases with different size of historical data}
\label{bayesian_ts_fig5}
\end{figure}

\begin{figure}
\centering
\includegraphics[width=0.65\linewidth]{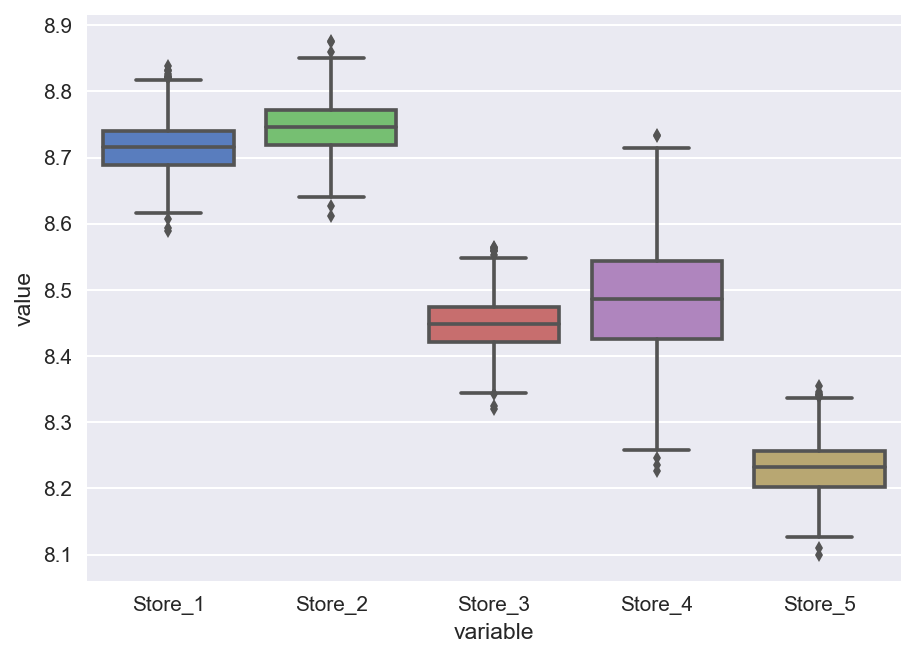}
\caption{Box plots for probability density function of intersect parameters of diferent time series in case of short time period of historical data for a specified store \textit{Store\_4}. }
\label{bayesian_ts_fig6}
\end{figure}

Let us consider Bayersian hierarchical model for time series. Intersect parameters $\alpha(Store)$ in the model (\ref{eq_hierarchical_model_1}) are different for different stores.
We have considered a case  with five different stores. 
Figure~\ref{bayesian_ts_fig4} shows the boxplots for the probability density functions  of intersect parameters. 
The dispersion of these dsitributions describes an uncertainty of influence of specified store on sales. 
 Hierarchical approach makes it possible to use such a model in the case with short historical data for specific stores, e.g. in the case of new stores. 
Figure~\ref{bayesian_ts_fig5}  shows the results of the forecasting on the validation set in the two cases, the fisrt case  is when we use 2 year historical data and the second case is with historical data for 5 days. We can see that such short historical data allow us to estimate sales dynamics correctly. Figure ~\ref{bayesian_ts_fig6}
shows box plots for probability density function of intersect parameters of different time series in case of short time period of historical data for a specified store \textit{Store\_4}. The obtained results show 
 that the dispersion for a specific store with short historical data becomes larger due to uncertainty caused by very short historical data for the specified store.

\begin{figure}
\centering
\includegraphics[width=0.85\linewidth]{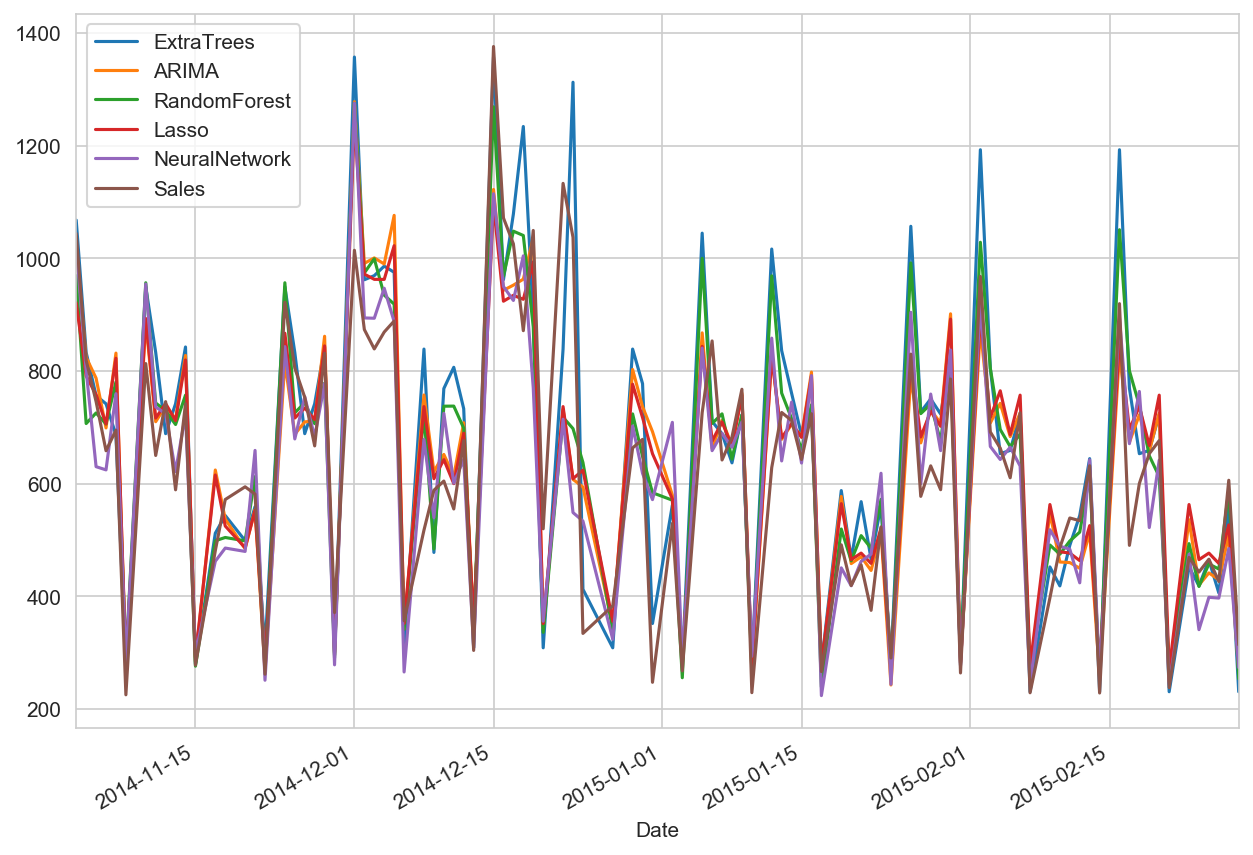}
\caption{Forecasting of different models on the validation set}
\label{fig1}
\end{figure}

Let us consider the results of Bayesian regression approach for stacking predictive models.
 We trained different predictive models and made the predictions on the validation set. 
 The  ARIMA model was evaluated using \textit{statsmodels} package, Neural Network was evaluated using \textit{keras} package, Random Forest and  Extra Tree was evaluated using \textit{sklearn} package. In these calculations, we used the approaches described in 
 ~\cite{pavlyshenko2019machine}.
Figure ~\ref{fig1} shows the time series forecasts on the validation sets obtained using different models. 

The results of prediction of these models on the validation sets are considered as
the  covariates for the regression on the second stacking level of the  ensemble of models.
For stacking predictive models we split the validation set on the training and testing sets.
For stacking regression, we normalized the covariates and target variable using 
z-scores:
\begin{equation}
z_{i}= \frac{x_{i}-\mu_{i}}{\sigma_{i}},
\end{equation}
where $\mu_{i}$ is the  mean value, $\sigma_{i}$ is the standard deviation. 
The prior distributions  for parameters $\alpha, \beta, \sigma $ in the  Bayesian  regression model 
(\ref{eq_1})-(\ref{eq_2})  are considered as Gaussian with mean values equal to 0, and standard deviation equal to 1. 
We split the validation set on the training and testing sets by time factor.
The parameters for prior distributions can be adjusted 
using prediction scores on testing sets or using expert opinions in the case of small 
data amount. 
To estimate uncertainty of regression coefficients, 
we used  the coefficient of variation
which is defined as a ratio between the 
standard deviation and mean value for model coefficient distributions:
\begin{equation}
v_{i}= \frac{\sigma_{i}}{\mu_{i}},
\end{equation}
where $v_{i}$ is the coefficient of variation, $\sigma_{i}$ is a standard deviation, $\mu_{i}$ is the mean value for the distribution of the regression coefficient of the model $i$.
 Taking into account that $\mu_{i}$ can be negative, we will analyze the absolute value of the coefficient of variation $\vert v_{i} \vert$.
 For the results evaluations, we used a relative mean absolute error (RMAE) and root mean square error (RMSE).  Relative mean absolute error (RMAE) was considered as a ratio between the mean absolute error (MAE) and mean values of target variable:
  \begin{equation}
RMAE=\frac{E ( \vert y_{pred} - y \vert  )} {E( y )}  100\%
\end{equation}
 Root mean square error (RMSE) was considered as:
  \begin{equation}
RMSE=\sqrt{{\frac{\sum_{i}^{n}(y_{pred}-y)^2}{n}}}
\end{equation}

\begin{figure}
\center
\includegraphics[width=0.75\linewidth]{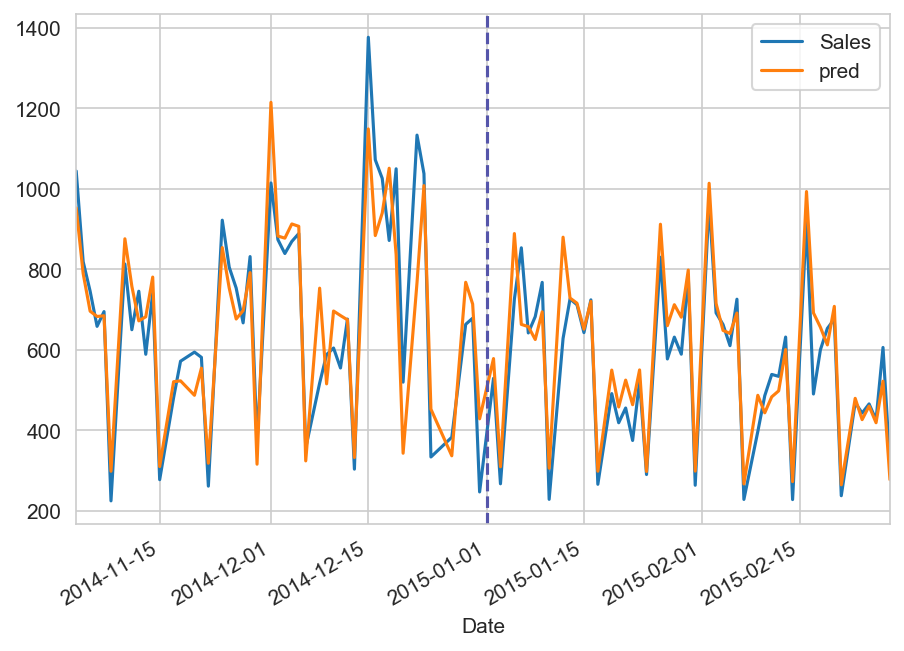}
\caption{Time series of mean values for real and forecasted sales on validation and testing sets}
\label{fig2}
\end{figure}

\begin{figure}
\centering
\includegraphics[width=0.65\linewidth]{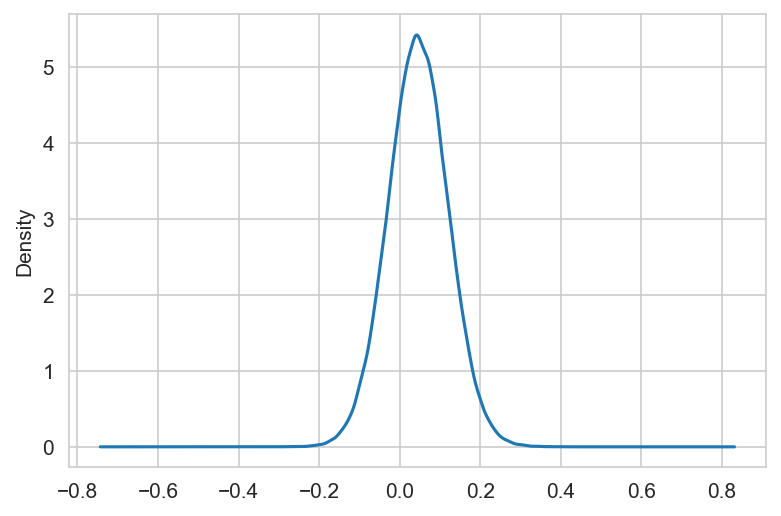}
\caption{The PDF for intersect parameter of stacking regression}
\label{fig3}
\end{figure}

 The data with predictions of different models  on  the validation set were split on the training set (48 samples) and testing set (50 samples) by date. 
We used the robust regression with Student's t-distribution for the target variable. 
As a result of calculations, we received the following scores: RMAE(train)=12.4\%, RMAE(test)=9.8\%, RMSE(train)=113.7, RMSE(test)= 74.7.
Figure ~\ref{fig2}  shows mean values time series for real and forecasted sales on the validation and testing sets.
The vertical dotted line  separates the training and testing sets.
Figure ~\ref{fig3}  shows the  probability density function (PDF) for the intersect parameter. One can observe 
a positive bias of this (PDF). It is caused by the fact that we applied machine learning  algorithms to  nonstationary time series. If a nonstationary trend is small, it can be compensated on  the validation set using stacking regression.  

\begin{figure}
\centering
\includegraphics[width=0.75\linewidth]{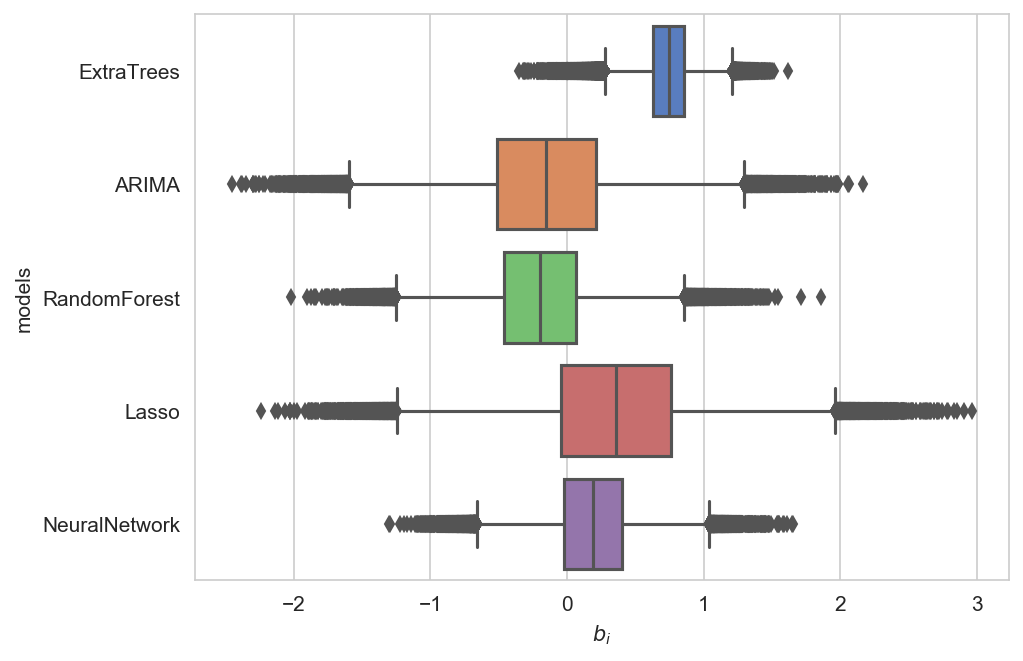}
\caption{Box plots for the PDF of  regression coefficients of models}
\label{fig4}
\end{figure}

\begin{figure}
\centering
\includegraphics[width=0.75\linewidth]{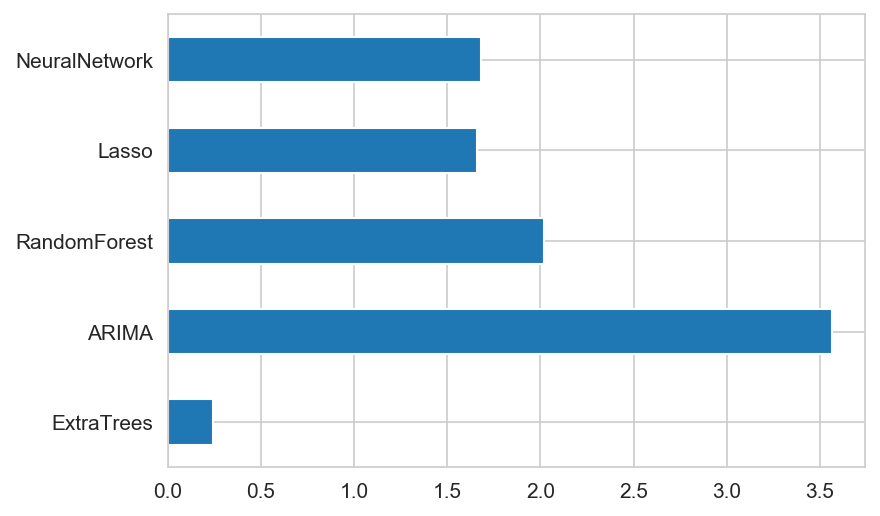}
\caption{Absolute values of the coefficient of variation for the PDF of  regression coefficients of models}
\label{fig5}
\end{figure}

Figure ~\ref{fig4}  shows the box plots for the PDF of  coefficients of models.
Figure ~\ref{fig5}  shows the coefficient of variation for the PDF of  regression coefficients of models. 

\begin{figure}
\centering
\includegraphics[width=0.75\linewidth]{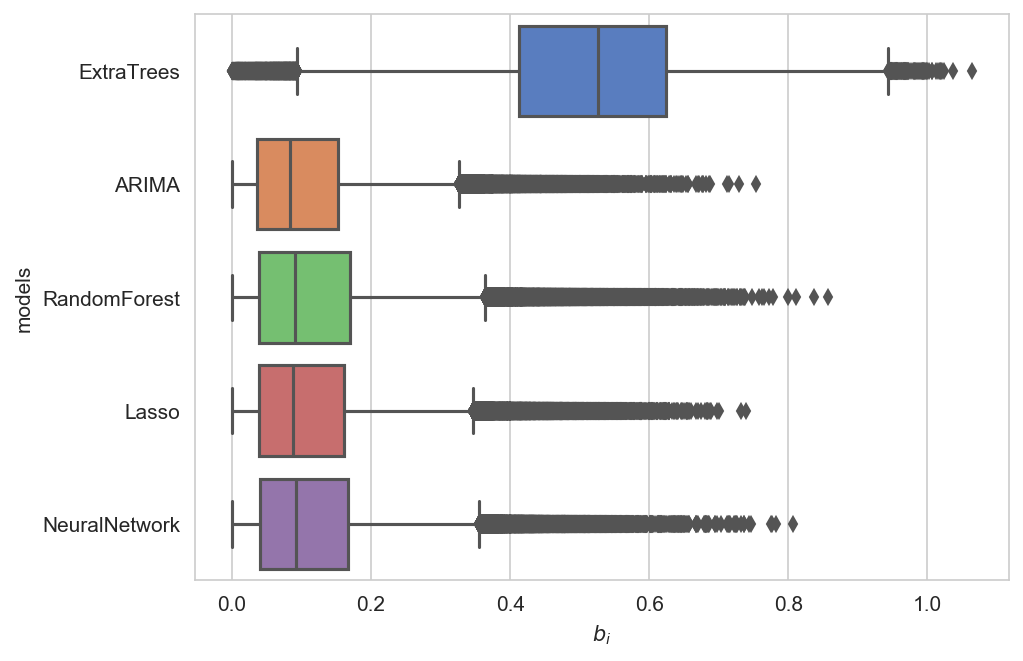}
\caption{Box plots for the PDF of  regression coefficients of models}
\label{fig6}
\end{figure}

We considered the case with the restraints to  regression coefficient of models that they should be positive.
We received the similar results:
RMAE(train)=12.9\%, RMAE(test)=9.7\%, RMSE(train)=117.3, RMSE(test)=76.1.
Figure ~\ref{fig6}  shows the box plots for the PDF of model regression  coefficients for this case.

\begin{figure}
\centering
\includegraphics[width=0.75\linewidth]{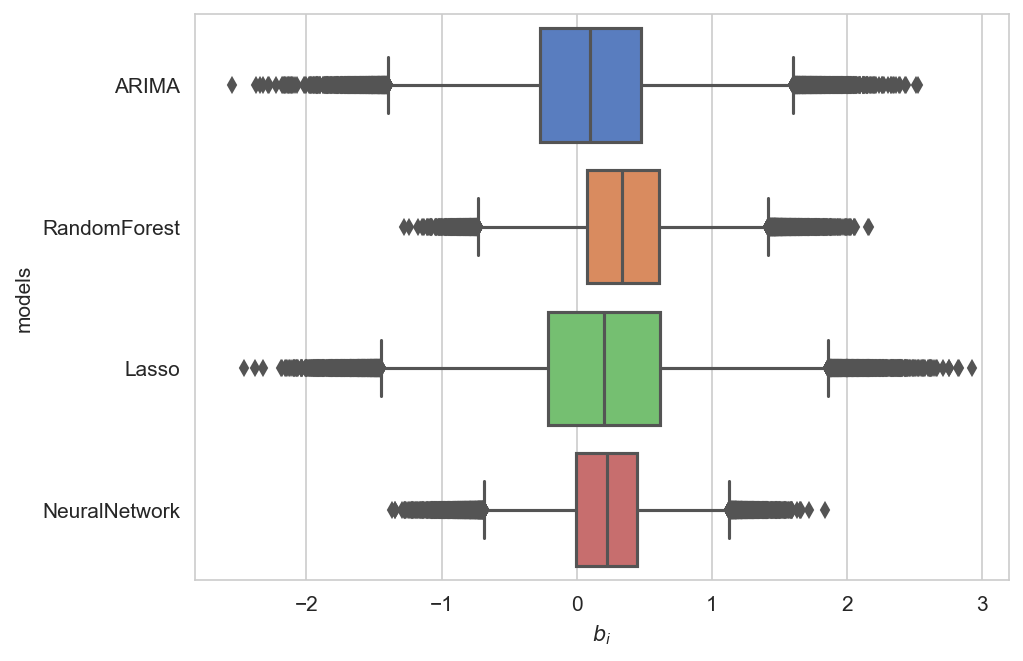}
\caption{Box plots for the PDF of  regression coefficients of models}
\label{fig7}
\end{figure}

\begin{figure}
\centering
\includegraphics[width=0.75\linewidth]{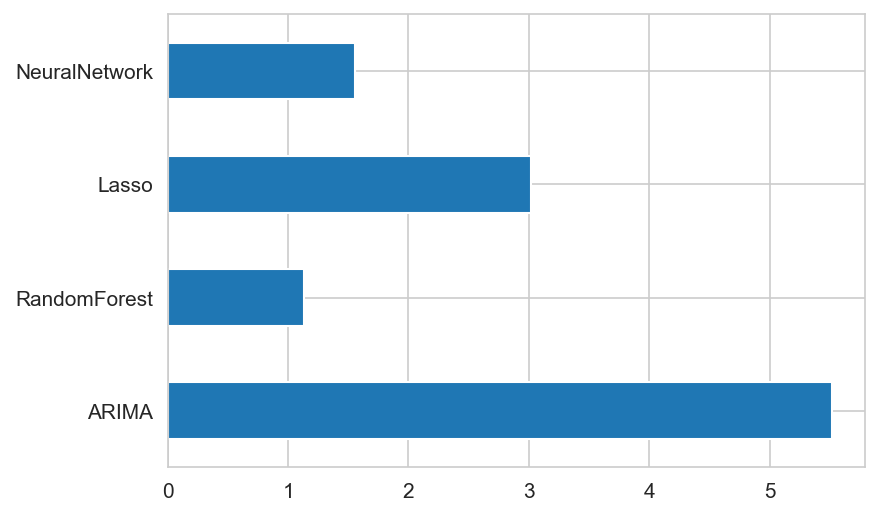}
\caption{Absolute values of the coefficient of variation for the PDF of models regression coefficients}
\label{fig8}
\end{figure}

\begin{figure}
\centering
\includegraphics[width=0.75\linewidth]{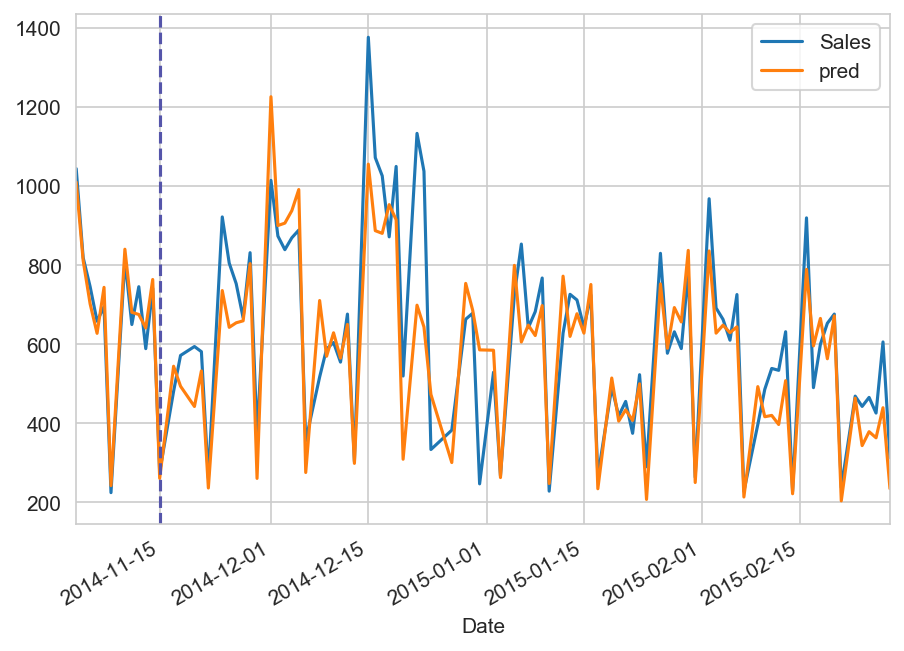}
\caption{Mean values time series for real and forecasted sales on the validation and testing sets in the case of small training set}
\label{fig9}
\end{figure}

\begin{figure}
\centering
\includegraphics[width=0.75\linewidth]{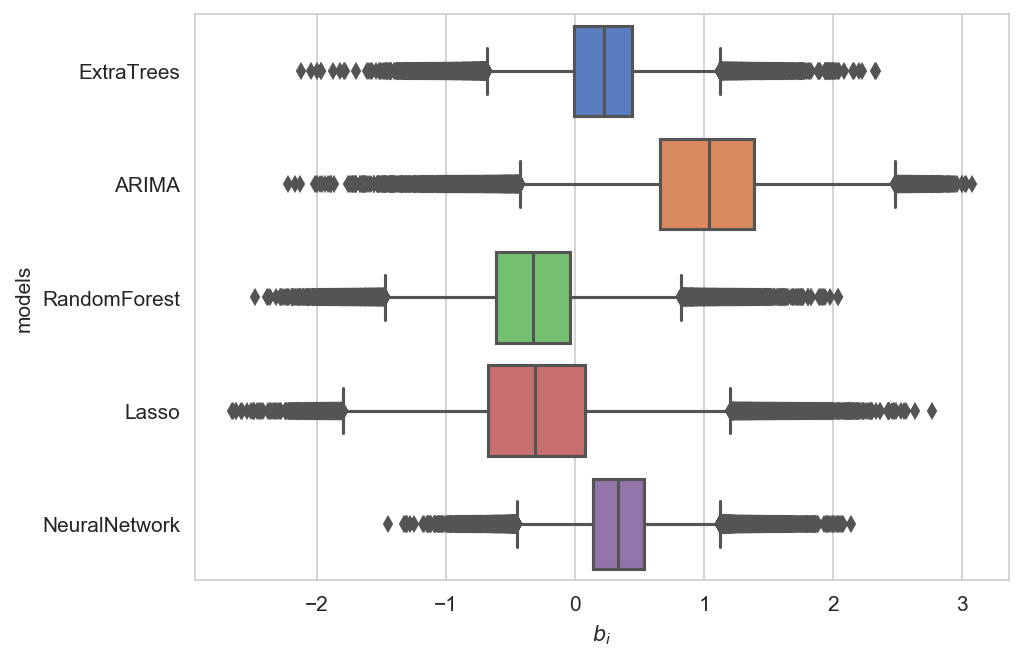}
\caption{Box plots for the PDF of regression coefficients of models in the case of small training set}
\label{fig10}
\end{figure}

\begin{figure}
\centering
\includegraphics[width=0.75\linewidth]{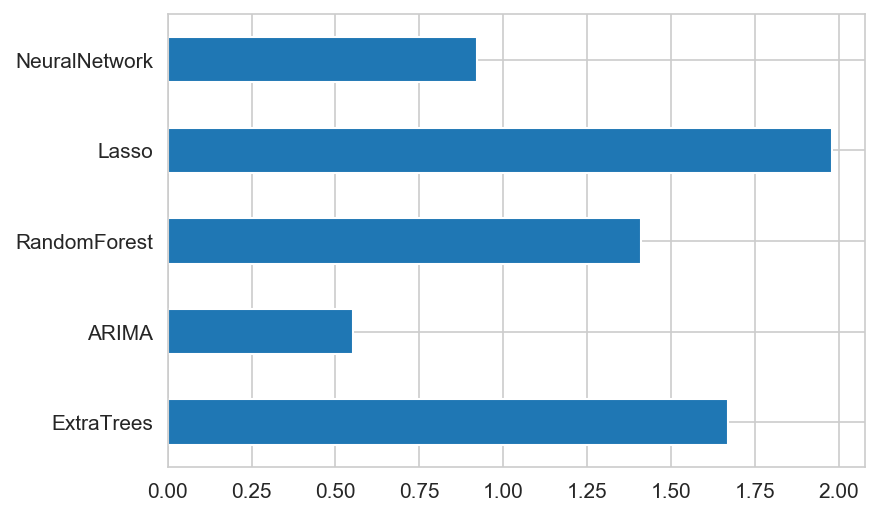}
\caption{Absolute values of the coefficient of variation for the PDF of  regression coefficients of models in the case of small training set}
\label{fig11}
\end{figure}

  All models have a similar mean value and variation coefficients. 
We can observe that errors characteristics RMAE and RMSE  on the validation set can be similar with respect to these errors on the training set. It tells us about the fact that Bayesian regression does not overfit on the training set comparing to the machine learning algorithms which can  demonstrate essential overfitting on  training sets, especially in the cases 
 of small amount of training data. 
We have chosen the best ExtraTree stacking model    and conducted Bayesian regression  with this one model only. We received the following scores:
RMAE(train)=12.9\%, RMAE(test)=11.1\%, RMSE(train)=117.1, RMSE(test)=84.7.
We also tried  to exclude the best model ExtraTree from the stacking regression and 
conducted Bayesian regression  with the rest  of models without ExtraTree. In this case
we received  the following scores:
RMAE(train)=14.1\%, RMAE(test)=10.2\%, RMSE(train)=139.1, RMSE(test)=75.3.
Figure~\ref{fig7}  shows the box plots for the PDF of model regression coefficients,
 figure~\ref{fig8}  shows the coefficient of variation for the PDF of regression coefficients of models for this case study.   We received worse results on the testing set. At the same time these models have the similar influence  and thus they  can potentially provide more stable results in the future  due to possible  changing of the quality of features. Noisy models can decrease accuracy on large training data sets, at the same time they contribute to sufficient results in the case of small data sets. 
We considered the case with a small number of training data, 12 samples. To get stable results, we fixed the $\nu$ parameter of Student's t-distribution in Bayesian regression model (\ref{eq_1})-(\ref{eq_2}) equal to 10.  
 We received  the following scores: RMAE(train)=5.0\%, RMAE(test)=14.2\%, RMSE(train)=37.5, 
 RMSE(test)=121.3.
Figure ~\ref{fig9}  shows mean values time series for real and forecasted sales on the validation and testing sets.  
Figure ~\ref{fig10}  shows the box plots for the PDF of regression coefficients of models.
Figure ~\ref{fig11}  shows the coefficient of variation for the PDF of  regression coefficients of models. In this case, we can see  that an other model starts playing an important role comparing with the previous cases and ExtraTree model does not dominate. 
The obtained results show that optimizing  informative prior distributions 
of stacking model parameters can improve the scores of prediction results on the testing set.

\section{Conclusion}
Bayesian regression approach for time series modeling  with nonlinear trend was analyzed. 
Such  approach allows us to estimate an uncertainty of time series prediction and  calculate value at risk (VaR) characteristics. 
Hierarchical model for time series using Bayesian regression has been considered. 
In this approach, one set of parameters is the same for all data samples, other parameters can be different for different groups of data samples. This approach 
 makes it possible to use such a model in the case with short historical data for specified time series, e.g. in the case of new stores or new products in the sales prediction problem.
 The use of Bayesian inference for time series predictive models stacking  has been analyzed.
A two-level ensemble of the predictive models for time series was considered.  
The models  ARIMA, Neural Network, Random Forest, Extra Tree were used for the prediction on the first level of ensemble of models. 
On the second stacking level, time series predictions of these models on the validation set were conducted by Bayesian regression. 
Such an approach gives distributions for regression coefficients of these models. It makes it possible to estimate the uncertainty  contributed by  each model to the stacking result.  
The information about these distributions allows us to select an optimal set of stacking models, taking into account domain knowledge.  Probabilistic approach for stacking predictive 
models allows us to make risk assessment for the predictions that is important in a decision-making process. 
Noisy models can decrease accuracy on large training data sets, at the same time they contribute to sufficient results in the case of small data sets. 
Using Bayesian inference for stacking  regression can be useful in cases of small datasets and help experts to select a set of models for stacking as well as make assessments of different kinds of risks. 
Choosing the final models for stacking is up to an expert  who takes into account different factors such as  uncertainty of each model on the stacking regression level,
   amount of training and testing data, the stability of models.
   In Bayesian regression,  we can receive a quantitative measure for the uncertainty that can be a very  useful information for experts in model selection and stacking. 
  An  expert can also set up informative prior distributions for  stacking regression  coefficients of models,
   taking into account the domain knowledge information. 
 So, Bayesian approach for stacking regression can give us the information about 
uncertainty of predictive models. Using this information and domain knowledge, 
an expert can select models to get stable stacking ensemble of predictive models.   

\bibliographystyle{splncs04}
\bibliography{article.bib}
\end{document}